\def\onecol{1}

\if\onecol1
\documentclass[apj,onecolumn]{openjournal}
\else
\documentclass[apj,twocolumn]{openjournal}
\fi

\usepackage{amsmath}
\graphicspath{{./}{figures/}}
\usepackage{xcolor}
\definecolor{xlinkcolor}{cmyk}{1.0, 0.31, 0, 0.42}

\PassOptionsToPackage{hyphens}{url}

\newcommand{\orcidauthor}[3]{\author{\href{http://orcid.org/#1}{#2$^{#3}$}}}

\usepackage[bookmarks=true, pdfnewwindow=true, colorlinks=true, linkcolor=xlinkcolor, citecolor=xlinkcolor, filecolor=xlinkcolor, urlcolor=xlinkcolor, final=true]{hyperref}
\usepackage{bbm}
\shorttitle{Testing Methods for Viscous Hydrodynamics}
\shortauthors{A. J. Dittmann \& G. Ryan }

\begin{document}

\title{\vspace{-0.8cm} A More Rigorous Test Problem For Viscous Hydrodynamics Codes\vspace{-1.8cm}}

\orcidauthor{0000-0001-6157-6722}{Alexander J. Dittmann}{1,*,\dagger}
\orcidauthor{0000-0001-9068-7157}{Geoffrey Ryan}{2}

\affiliation{$^{1}$ Institute for Advanced Study, 1 Einstein Drive, Princeton, NJ 08540, USA}
\affiliation{$^{2}$ Perimeter Institute for Theoretical Physics, 31 Caroline St. N., Waterloo, ON, N2L 2Y5, Canada}

\thanks{$^*$\href{mailto:dittmann@ias.edu}{dittmann@ias.edu}}
\thanks{$^{\dagger}$ NASA Einstein Fellow}

\begin{abstract}
We advocate for a more stringent test problem for codes that aim to solve the equations of viscous hydrodynamics. Specifically, we discuss a nonuniform-density version of the common (uniform-density) Gaussian velocity shear test, where density gradients transverse to the direction of velocity shear cause the velocity profile to drift over time. By employing a nonunifom density, this test provides a test that the full viscous stress (and velocity shear) tensors are calculated correctly from the conserved variables, and checks the correctness of the fluxes and source terms calculated therefrom.

In Appendix \ref{app:equations}, we present a detailed exposition of the Navier Stokes equations, particularly their fluxes and source terms in a variety of common coordinate systems.
\end{abstract}

\section{Introduction}\label{sec:intro}
A standard problem used to test codes designed to solve the Navier-Stokes equations is following the evolution of a Gaussian velocity bump \citep[e.g.,][]{2013MNRAS.436.2997D,2016ApJS..226....2D,2024arXiv240916053S}, inspired by analogy to the classic solution to the heat equation. However, this test is typically employed using a constant background density, which can allow errors to slip into the computation of viscous evolutionary terms undetected or underappreciated. In the following we discuss a more general analytical solution to the equations of viscous hydrodynamics and its application as a test problem for numerical schemes. 

\section{A Cartesian Velocity Shear Test with Nonuniform Background Density}
We consider a one-dimensional shear flow, uniform in the $y$ and $z$ directions. We assume that the fluid pressure is related simply to the density and a spatially-dependent sound speed, $P=c_s^2(x)\rho$, that the fluid is static in the $z$-direction, and that the bulk viscosity of the fluid is zero. 
The corresponding equations of compressible viscous hydrodynamics written in Cartesian coordinates are
\begin{align}
\partial_t\rho + \partial_x(\rho v_x) &= 0, \\
\partial_t(\rho v_x) + \partial_x(\rho v_x^2 + P - 2\eta\sigma_{xx}) &= 0, \\
\partial_t(\rho v_y) + \partial_x(\rho v_xv_y -2 \eta\sigma_{xy}) &= 0,
\end{align}
where $\eta$ is the dynamic viscosity (related to the kinematic viscosity by $\eta\equiv\rho\nu$) and $\sigma_{ij}$ are components of the velocity shear tensor. (See Appendix \ref{app:equations} for more general expressions in various coordinate systems.) The velocity shear tensor is symmetric and trace-free in $d$ dimensions, and given in Cartesian coordinates by 
\begin{equation}
\sigma_{ij} = \frac{1}{2}\left(\partial_i v_j + \partial_j v_i\right) - \gamma_{ij}\frac{1}{d}\nabla\cdot v,
\end{equation}
where $\gamma$ is the metric tensor, in Cartesian coordinates the unit dyadic. 
The relevant terms of the shear tensor are thus given by
\begin{align}
\sigma_{xx}&=\partial_xv_x - \frac{1}{d}\partial_xv_x = \frac{d-1}{d}\partial_xv_x,\\
\sigma_{xy}&=\frac{1}{2}\partial_xv_y.
\end{align}

Using the continuity equation, the momentum equations can be written as
\begin{align}
\rho\partial_tv_x + \rho v_x \partial_xv_x + \partial_x(c_s^2\rho) - 2\partial_x(\eta\sigma_{xx})=0\\
\rho\partial_tv_y + \rho v_x\partial_xv_y - 2\partial_x(\eta\sigma_{xy})=0.
\end{align}

We note that so long as the pressure remains uniform in $x$, $P(x) = P_0$ (i.e. $\partial_x (c_s^2\rho) = 0$) then an initially $x$-static $v_x = 0$ will remain so for all time. Consequentially $\partial_t \rho = 0,$ and therefore the density distribution remains fixed in time. Trivially, a uniform $v_x$ may be added which will cause the fluid pattern to simply advect.

Under the aforementioned assumptions, the evolution of the fluid in the $y$ direction is simply
\begin{equation}
\rho\partial_tv_y=\partial_x(2\eta\sigma_{xy})=\partial_x(\eta\partial_xv_y).
\end{equation}
To simplify things further, we assume a simple density profile: $\rho(x)=\rho_0\exp{(-\kappa x)}$. Under this assumption and that of a globally-constant kinematic viscosity, the evolution of $v_y$ reduces to

\begin{equation}\label{eq:theProblem}
\partial_tv_y+\nu\kappa\partial_xv_y-\nu\partial^2_xv_y=0. 
\end{equation}

A solution to Equation (\ref{eq:theProblem}) for $t>0$ is
\begin{equation}
v_y(x,t)=\frac{A}{\sqrt{4\pi\nu t}}\exp{\left(-\frac{(x-\kappa\nu t)^2}{4\nu t}\right)},
\end{equation}
with an arbitrary initial amplitude scale $A$. 
Thus, an initial Gaussian shear profile will spread over time, with its variance evolving as $\sigma^2=2\nu t$ and the center shifting along the density gradient with velocity $\nu\kappa$. Setting $\kappa=0$ reduces this to the standard velocity shear test used previously in the literature.

Although this test is quite simple analytically, attempting to calculate the solution to this problem in an inconvenient coordinate system (for example, orienting the shear and density profiles along directions misaligned with coordinate axes or solving for the evolution of this Cartesian problem using polar coordinates) can provide a nontrivial test of every component in the viscous stress tensor. 

It is worth noting that in the preceding derivation we have ignored the energy equation. In principle, viscous dissipation may heat the fluid and introduce pressure gradients and thus complicate the evolution of the velocity field. Thus, in practice, this is a useful test when employed alongside rapid cooling to maintain $\partial_x(c_s^2\rho)=0$, or an isothermal equation of state fixing $c_s\ll v$. The latter strategy, which we adopt in the following, sets a floor to the errors achievable when using this problem to test the convergence of a particular code. 

\section{Examples}\label{sec:nsmn}
In the following we apply three different codes to this problem, testing the accuracy of the codes and highlighting the ability of this test to identify errors passed over by the constant-density version of this test. 

We construct an initial state by choosing an initial standard deviation $\sigma_0$ and amplitude $v_0$. The former, along with the kinematic viscosity, determine an effective initial time $t_0=\sigma_0^2(2\nu)$ and amplitude $A=v_0\sqrt{4\nu t_0}$. In the following, we have used $v_0=10^{-3}$, $\sigma_0=0.1$, $\nu=1$. We set the initial location of the velocity Gaussian to $x_0=0.5$ and ran each simulation until $t=0.02$. As an example, we have conducted a series of tests using both uniform and nonuniform backgrounds, specifically $\kappa=0$ and $\kappa=15$.

For the sake of demonstration, we have implemented this test problem in two different codes, \texttt{Athena++} \citep{2020ApJS..249....4S} and \texttt{Disco}. Our \texttt{Athena++} simulations employed both Cartesian coordinates (with a uniform mesh) and cylindrical polar coordinates (with a log-uniform radial mesh spacing). Our \texttt{Disco} simulations were all conducted in cylindrical polar coordinates but used a grid linearly spaced in radius and with a variable number of cells in the $\phi$ direction to achieve nearly-equal-aspect ratio cells; in \texttt{Disco} we used both the approximate viscous scheme described in \citet{2016ApJS..226....2D} (``\texttt{v2016}'') and a complete shear viscosity as described in \citet{2021ApJ...921...71D}.\footnote{Specifically, the \texttt{Disco v2016} viscosity implementation \textit{assumes} that the dynamic viscosity is globally uniform, which has led to a variety of erroneous results when the code has been used to simulate systems such as circumbinary disks \citep[see, for example, Appendix A of][for further discussion]{2024ApJ...967...12D}. This limitation also applies to simulations of gap-opening planets, for example.} 
In each case we used second-order Runge-Kutta time-stepping and second-order piecewise-linear reconstruction. 

\begin{figure}
\if\onecol1
\includegraphics[width=\linewidth]{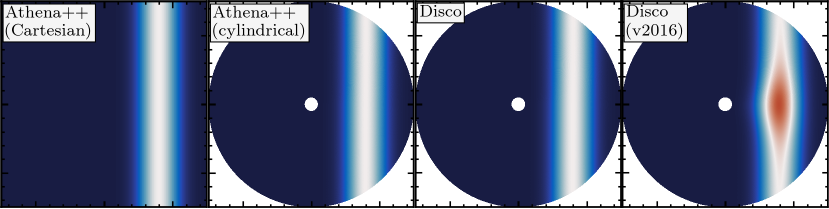}
\else
\includegraphics[width=\linewidth]{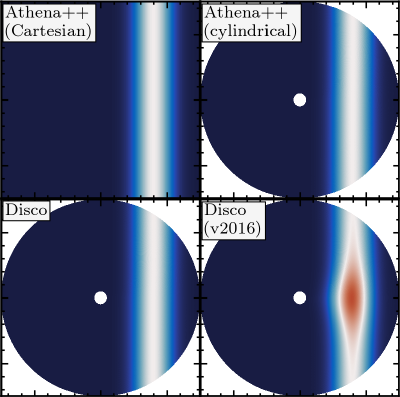}
\fi
\caption{Snapshots of $v_y$ at $t=0.02$ for simulations of a $\kappa=15$ test problem. The color scale is the same in each panel, such that red colors indicate velocities in excess of the maximum of the analytical solution.}\label{fig:vyprof}
\end{figure}

\begin{figure}
\if\onecol1
\includegraphics[width=\linewidth]{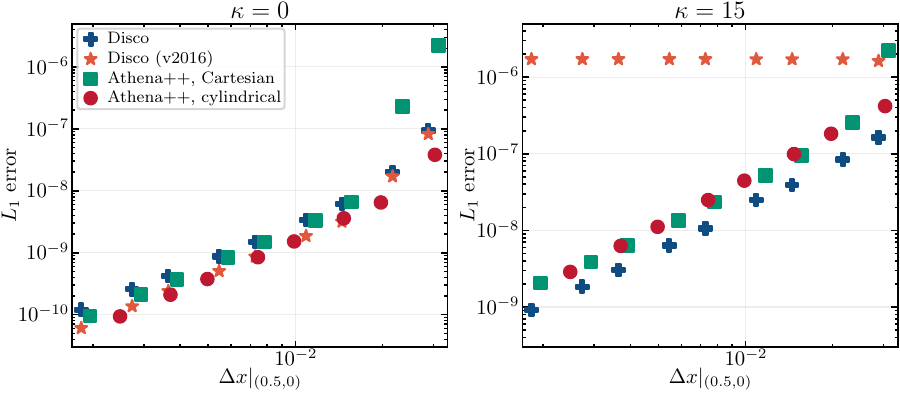}
\else
\includegraphics[width=\linewidth]{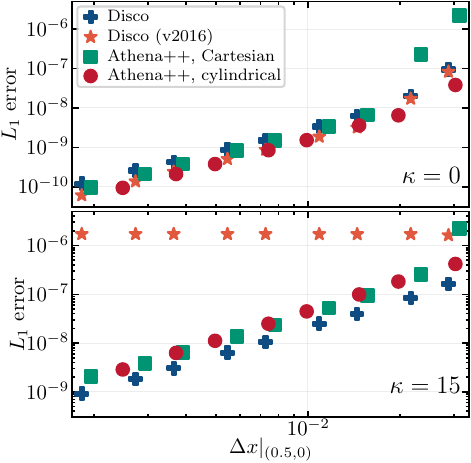}
\fi
\caption{The $L_1$ error of the $y-$velocity in each simulation, as a function of the cell size at $x=0,5,\,y=0$ (note that while the \texttt{Disco} and Cartesian simulations used uniformly sized cells, the cylindrical \texttt{Athena++} simulations had $\Delta r \propto r\Delta \phi \propto r$). At $\kappa=0$, each code converges at second order. However, for $\kappa=15$, \texttt{Disco v2016} fails to converge to the analytical solution.}
\label{fig:converge}
\end{figure}

Figure \ref{fig:vyprof} illustrates the $v_y$ profile for a variety of simulations at $t=0.02$ for simulations employing $\kappa=15$ density profiles. As far as the eye can tell, \texttt{Athena++} and \texttt{Disco} both agree quite well with each other (and the analytical solution, as the center of the velocity bumps have shifted from $x=0.5$ to $x=0.8$), while \texttt{Disco v2016} deviates significantly. 

Figure \ref{fig:converge} illustrates the convergence, or lack thereof, of each code for both the $\kappa=0$ and $\kappa=15$ versions of this test problem. Notably, each code performs adequately with a uniform-density-background, but the limitations of \texttt{Disco v2016} are revealed at $\kappa\neq 0$. Only testing the $\kappa=0$ case might leave one unduly confident in the accuracy of one's code, and thus $\kappa\gg1$ should be preferred. 

\section{Summary}
We hope that this more rigorous, ``improved'' even, Gaussian velocity shear problem will prove useful for testing future numerical schemes. As illustrated above, it is more discriminating that the uniform-density version, in addition to being an intrinsically more thorough test. 

\section*{Acknowledgments}
AJD was funded by the NASA Hubble Fellowship grant
No. HST-HF2-51553.001 awarded by the Space Telescope Science Institute, which is operated by the Association of Universities for Research in Astronomy, Inc., for NASA, under contract NAS5-26555.
Research at Perimeter Institute is supported in part by the Government of Canada through the Department of Innovation, Science and Economic Development and by the Province of Ontario through the Ministry of Colleges and Universities. We thank Paul Duffell for his comments on this note, and for suggesting the appendix.

\vspace{6cm}
\bibliographystyle{aasjournal}
\bibliography{references}

\pagebreak
\appendix

\section{Compressible Navier-Stokes In Curvilinear Coordinates}\label{app:equations}

Here we give a brief derivation of the Navier-Stokes equations in curvilinear coordinates, with a particular eye for the viscous flux and source terms required for a numerical code like \texttt{Disco}.

The momentum and energy equations for a single fluid of density $\rho$, velocity $\bf{v}$, and specific internal energy $\epsilon$ with stress tensor ${\bf T}$ and under a gravitational field ${\bf g}$ are:
\begin{align}
    \partial_t \left( \rho \right)  + {\bf \nabla} \cdot \left( \rho {\bf v}\right) &= 0 \\
    \partial_t \left(\rho {\bf v}\right) + {\bf \nabla} \cdot \left ( \rho {\bf v} \otimes {\bf v} + {\bf T} \right) &= \rho {\bf g} \\
    \partial_t \left(\tfrac{1}{2}\rho v^2 + \rho \epsilon \right) + {\bf \nabla} \cdot \left ( \tfrac{1}{2}\rho v^2 {\bf v} + \rho \epsilon {\bf v} + {\bf T}\cdot {\bf v} \right) &= \rho {\bf v} \cdot {\bf g}
\end{align}

In the Navier-Stokes equations, the stress tensor includes isotropic pressure $P$ as well as a viscous stress ${\bf \tau}$: ${\bf T} = P{\bf 1} - {\bf \tau}$.

A curvilinear coordinate system is defined by its metric tensor $\gamma_{ij}$, where $i,j \in \{1,2,3\}$.  Vectors and tensors can be expressed as contravariant components $v^i$ or covariant components $v_i$ in the coordinate basis, or in an orthonormal basis $v^{\hat{i}} = v_{\hat{i}}$. In a cylindrical polar coordinate system, $v^\phi$ is the angular velocity, $v_\phi$ the specific angular momentum, and $v^{\hat{\phi}}$ the component of linear velocity in the $\phi$ direction.  In coordinate form the Navier-Stokes system can be written:
\begin{align}
    \partial_t \left( \rho \right)  + \nabla_j \left( \rho  v^j\right) &= 0 ,\\
    \partial_t \left(\rho v_i\right) + \nabla_j \left ( \rho v_i v^j + T^j_i \right) &= \rho g_i ,\\
    \partial_t \left(\tfrac{1}{2}\rho v^2 + \rho \epsilon \right) + \nabla_j \left ( \tfrac{1}{2}\rho v^2 v^j+ \rho \epsilon v^j + T^j_k v^k \right) &= \rho v^j g_j,
\end{align}
where $\nabla_i$ is the covariant derivative for metric tensor $\gamma_{ij}$.  Numerical algorithms require us to express these equations in terms of partial derivatives $\partial_i$ instead of covariant derivatives.  Two identities help us, for a vector {\bf v} and symmetric tensor ${\bf T}$:
\begin{align}
    \nabla_j v^j &= \frac{1}{\sqrt{\gamma}} \partial_j \left( \sqrt{\gamma} v^j \right),  \\
    \nabla_j T^j_i &= \frac{1}{\sqrt{\gamma}} \partial_j \left( \sqrt{\gamma} T^j_i \right) - \tfrac{1}{2} T^{jk} \partial_i \gamma_{jk},
\end{align}
where $\gamma = \det \gamma_{ij}$ is the determinant of the metric tensor. The explicit coordinate form of the compressible Navier-Stokes system can then be expressed:
\begin{align}
    \partial_t \left( \rho \right)  + \frac{1}{\sqrt{\gamma}}\partial_j \left( \sqrt{\gamma} \rho  v^j\right) &= 0 ,\\
    \partial_t \left(\rho v_i\right) + \frac{1}{\sqrt{\gamma}}\partial_j \left[\sqrt{\gamma} \left ( \rho v_i v^j + T^j_i \right)\right] &= \tfrac{1}{2}\left(\rho v^j v^k + T^{jk}\right) \partial_i \gamma_{jk} + \rho g_i ,\\
    \partial_t \left(\tfrac{1}{2}\rho v^2 + \rho \epsilon \right) + \frac{1}{\sqrt{\gamma}}\partial_j \left[\sqrt{\gamma} \left ( \tfrac{1}{2}\rho v^2 v^j+ \rho \epsilon v^j + T^j_k v^k \right)\right] &= \rho v^j g_j.
\end{align}

A generic conservation law for quantity $U$ with flux ${\bf F}$ and source $S$ takes the form:
\begin{align}
    \partial_t U + \frac{1}{\sqrt{\gamma}} \partial_j \left(\sqrt{\gamma} F^j \right) = S. \label{eq:consLaw}
\end{align}

To construct a numerical scheme in the finite-volume formulation one considers a control volume $V$ with boundary $\partial V$. The boundary has a unit normal vector ${\bf n}$ (ie. ${\bf n} \cdot {\bf n} = 1$), a two dimensional metric tensor $\Sigma_{ab}$, and area element $\sqrt{\Sigma} d^2x$, where $\Sigma = \det{\Sigma_{ab}}$.  Integrating Equation \eqref{eq:consLaw} over $V$ and applying Stokes' Theorem provides an equation for the evolution of the mass of $U$ in $V$:
\begin{align}
    \frac{d}{dt} \int_V U \sqrt{\gamma}d^3x = -\int_{\partial V} F^j n_j \sqrt{\Sigma}d^2x + \int_V S\sqrt{\gamma}d^3x .
\end{align}

A finite volume numerical scheme approximately solves this equation by discretizing the integrals, determining an approximate numerical flux, and integrating in time.  A simple schematic of this is:
\begin{align}
    \frac{d}{dt} U \Delta V \approx -\sum_{f \in \mathrm{faces}} F^{j}n^{(f)}_j \Delta A_f + S \Delta V,
\end{align}
where $\Delta V = \int_V \sqrt{\gamma}d^3x$ is the volume of $V$, and the boundary $\partial V$ is composed of a finite number of faces $f$ each with unit normal ${\bf n}^{(f)}$ and area $\Delta A_f = \int_{f} \sqrt{\Sigma}d^2x$.  The relevant flux for the numerical scheme is hence $F^j n_j^{(f)} = {\bf F} \cdot {\bf n}^{(f)}$ for each face, the inner product of the flux with the unit normal for the face.  If, as in many numerical schemes, the faces are aligned with the coordinate system (e.g. surfaces of constant $x$, $y$, $z$ in a Cartesian code, or surfaces of constant $r$, $\phi$, $z$ in a cylindrical polar code), then the relevant flux in direction $j$ is simply the orthonormal component: ${\bf F} \cdot {\bf n}^{(j)} = F^{\hat{j}}$.

In the Navier-Stokes system we have:
\begin{align}
    U &= \begin{pmatrix} 
    \rho\\
    \rho v_i\\
    \tfrac{1}{2}\rho v^2 + \rho \epsilon 
    \end{pmatrix} & F^{\hat{j}} &= \begin{pmatrix} 
    \rho v^{\hat{j}} \\
        \rho v_i v^{\hat{j}} + T_i^{\hat{j}} \\
        \left( \tfrac{1}{2}\rho v^2 + \rho \epsilon\right )v^{\hat{j}} + T^{\hat{j}}_k v^k
        \end{pmatrix} & S &= \begin{pmatrix} 
        0 \\
        \tfrac{1}{2} \left(\rho v^j v^k + T^{jk}\right) \partial_i \gamma_{jk} + \rho g_i\\
        \rho v^j g_j
        \end{pmatrix}
\end{align}

The stress tensor $T_{jk} = P \delta_{jk} - \tau_{jk}$ where $\tau_{jk}$ is the viscous stress tensor.  The viscous stress depends on symmetric parts of the velocity gradient, specifically the shear $\sigma_{jk}$ and divergence $\Theta$:
\begin{align}
    \sigma_{ij} &= \tfrac{1}{2}\left( \nabla_i v_j + \nabla_j v_i\right) - \tfrac{1}{3} \gamma_{ij} \nabla_k v^k & \Theta &= \nabla_k v^k.
\end{align}

In all coordinate systems the shear tensor is symmetric $\sigma_{ij} = \sigma_{ji}$ and traceless $\sigma^i_i = \gamma^{ij} \sigma_{ij} = \gamma_{ij} \sigma^{ij} = 0$.  The viscous stress tensor is:
\begin{align}
    \tau_{ij} &= 2 \eta \sigma_{ij} + \zeta \gamma_{ij} \Theta,
\end{align}
with dynamic viscosity $\eta \geq 0$ and bulk viscosity $\zeta \geq 0$ (the kinematic viscosity is $\nu = \eta / \rho$).

Explicitly, then, the fluxes and sources are:
\begin{align}
    F^{\hat{j}} &= \begin{pmatrix} 
    \rho v^{\hat{j}} \\
        \rho v_i v^{\hat{j}} + \left(P - \zeta \Theta\right)n_i^{(j)} - 2\eta \sigma_i^{\hat{j}} \\
        \left( \tfrac{1}{2}\rho v^2 + \rho \epsilon + P - \zeta \Theta\right )v^{\hat{j}} - 2\eta \sigma_k^{\hat{j}} v^k 
        \end{pmatrix} & S &= \begin{pmatrix} 
        0 \\
        \tfrac{1}{2} \left(\rho v^j v^k + \left(P - \zeta \Theta\right) \gamma^{jk} - 2\eta \sigma^{jk}\right) \partial_i \gamma_{jk} + \rho g_i\\
        \rho v^j g_j
        \end{pmatrix}
\end{align}

\section{Cartesian Coordinates}

In Cartesian coordinates $\gamma_{ij} = \mathrm{diag}(1, 1, 1)$.  Contravariant, covariant, and orthonormal tensor components are all equal: $v^x = v_x = v^{\hat{x}} = v_{\hat{x}}$.  The shear tensor and divergence are:
\begin{align}
    \sigma_{i\ }^{\ j} &= \begin{pmatrix}
        \partial_x v^x - \tfrac{1}{3}\Theta & \tfrac{1}{2}\left( \partial_x v^y + \partial_y v^x \right) & \tfrac{1}{2}\left( \partial_x v^z + \partial_z v^x \right)\\
        \tfrac{1}{2}\left( \partial_x v^y + \partial_y v^x \right) & \partial_y v^y - \tfrac{1}{3}\Theta & \tfrac{1}{2}\left( \partial_y v^z + \partial_z v^y \right)\\
        \tfrac{1}{2}\left( \partial_x v^z + \partial_z v^x \right) & \tfrac{1}{2}\left( \partial_y v^z + \partial_z v^y \right) & \partial_z v^z - \tfrac{1}{3}\Theta\ .
    \end{pmatrix} \\
    \Theta &= \partial_x v^x + \partial_y v^y + \partial_z v^z
\end{align}

The viscous energy flux depends on the shear tensor dotted with the velocity:
\begin{align}
    \left( {\bf v} \cdot {\bf \sigma}\right)^j =  v^k\sigma_k^{\ j} &= \begin{pmatrix}
        v^x \partial_x v^x + \tfrac{1}{2}v^y\left(\partial_y v^x +  \partial_x v^y\right) + \tfrac{1}{2}v^z \left(\partial_z v^x + \partial_x v^z \right) - \tfrac{1}{3}v^x \Theta \\
        \tfrac{1}{2}v^x\left( \partial_x v^y + \partial_y v^x  \right)
        + v^y \partial_y v^y 
        + \tfrac{1}{2}v^z\left( \partial_z v^y + \partial_y v^z \right) - \tfrac{1}{3}v^y\Theta \\
        \tfrac{1}{2}v^x\left( \partial_x v^z + \partial_z v^x \right) + \tfrac{1}{2}v^y\left( \partial_y v^z + \partial_z v^y \right) +  v^z\partial_z v^z - \tfrac{1}{3}v^z\Theta
        \end{pmatrix}
\end{align}

The conserved variables and sources are simply:
\begin{align}
     U &= \begin{pmatrix} 
    \rho \\
    \rho v^x \\
    \rho v^y \\
    \rho v^z \\
    \tfrac{1}{2}\rho v^2 + \rho \epsilon
    \end{pmatrix}, & S &= \begin{pmatrix} 
        0 \\
        \rho g_x\\
        \rho g_y\\
        \rho g_z\\
        \rho \left(v^x g_x + v^y g_y + v^z g_z\right)
        \end{pmatrix}.
\end{align}

The fluxes are only somewhat unwieldy when written explicitly:
\begin{align}
    F^{\hat{x}} &= \begin{pmatrix}
    \rho v^x \\
    \rho {v^x}^2 + P - 2\eta \partial_x v^x + \left(\tfrac{2}{3}\eta - \zeta\right)\Theta \\
    \rho v^x v^y - \eta \left(\partial_x v^y + \partial_y v^x\right)  \\
    \rho v^x v^z - \eta \left(\partial_x v^z + \partial_z v^x\right)  \\
    \left(\tfrac{1}{2}\rho v^2 + \rho \epsilon + P + \left(\tfrac{2}{3}\eta - \zeta\right)\Theta\right)v^x - 2 \eta \left ( {\bf v}\cdot \sigma \right)^x
    \end{pmatrix} \\
    F^{\hat{y}} &= \begin{pmatrix}
    \rho v^y \\
    \rho v^x v^y - \eta \left(\partial_x v^y + \partial_y v^x\right)  \\
    \rho {v^y}^2 + P - 2\eta \partial_y v^y + \left(\tfrac{2}{3}\eta - \zeta\right)\Theta \\
    \rho v^y v^z - \eta \left(\partial_y v^z + \partial_z v^y\right)  \\
    \left(\tfrac{1}{2}\rho v^2 + \rho \epsilon + P + \left(\tfrac{2}{3}\eta - \zeta\right)\Theta\right)v^y - 2\eta \left( {\bf v}\cdot \sigma\right)^y
    \end{pmatrix} \\
    F^{\hat{z}} &= \begin{pmatrix}
    \rho v^z \\
    \rho v^x v^z - \eta \left(\partial_x v^z + \partial_z v^x\right)  \\    
    \rho v^y v^z - \eta \left(\partial_y v^z + \partial_z v^y\right)  \\
    \rho {v^z}^2 + P - 2\eta \partial_z v^z + \left(\tfrac{2}{3}\eta - \zeta\right)\Theta \\
    \left(\tfrac{1}{2}\rho v^2 + \rho \epsilon + P + \left(\tfrac{2}{3}\eta - \zeta\right)\Theta\right)v^z - 2\eta \left({\bf v}\cdot \sigma \right)^z
    \end{pmatrix}
\end{align}

\section{Cylindrical Polar Coordinates}

Cylindrical polar coordinates $(r, \phi, z)$ have a metric tensor $
\gamma_{ij} = \mathrm{diag}(1, r^2, 1)$.  Radial and vertical tensor components are as in the Cartesian case, but azimuthal components obey $v_\phi = r^2 v^\phi$, $v^{\hat{\phi}} = v_{\hat{\phi}} = r v^\phi$.  We give expressions in terms of $v^r$, $v^\phi$, and $v^z$: the radial, angular (measured in e.g. [rad/s]), and vertical velocities.  The unit normals in each dimension are:
\begin{align}
    n^{(r)}_i &= \begin{pmatrix} 1 \\ 0 \\ 0 \end{pmatrix} & n^{(\phi)}_i &= \begin{pmatrix} 0 \\ r \\ 0 \end{pmatrix} & n^{(z)}_i = \begin{pmatrix} 0 \\ 0 \\ 1\end{pmatrix}
\end{align}

The divergence is:
\begin{align}
    \Theta = \nabla_k v^k = \partial_r v^r + \partial_\phi v^\phi + \partial_z v^z + \tfrac{1}{r} v^r .
\end{align}

The shear tensor in these coordinates is:
\begin{align}
    \sigma_{i\ }^{\ j} &= \begin{pmatrix}
        \partial_r v^r - \tfrac{1}{3}\Theta & \tfrac{1}{2}\left( \partial_r v^\phi + \tfrac{1}{r^2}\partial_\phi v^r \right) & \tfrac{1}{2}\left( \partial_r v^z + \partial_z v^r \right)\\
        \tfrac{1}{2}\left( \partial_\phi v^r + r^2 \partial_r v^\phi \right) & \partial_\phi v^\phi + \tfrac{1}{r}v^r - \tfrac{1}{3}\Theta & \tfrac{1}{2}\left( \partial_\phi v^z + r^2 \partial_z v^\phi \right)\\
        \tfrac{1}{2}\left( \partial_z v^r + \partial_r v^z \right) & \tfrac{1}{2}\left( \partial_z v^\phi + \tfrac{1}{r^2}\partial_\phi v^z \right) & \partial_z v^z - \tfrac{1}{3}\Theta
    \end{pmatrix}.
\end{align}

Dotted with velocity (for the energy flux):
\begin{align}
    \left( {\bf v} \cdot {\bf \sigma}\right)^j =  v^k\sigma_k^{\ j} &= \begin{pmatrix}
        v^r \partial_r v^r + \tfrac{1}{2}v^\phi\left(\partial_\phi v^r + r^2 \partial_r v^\phi\right) + \tfrac{1}{2}v^z \left(\partial_z v^r + \partial_r v^z \right) - \tfrac{1}{3}v^r \Theta \\
        \tfrac{1}{2}v^r\left( \partial_r v^\phi + \tfrac{1}{r^2}\partial_\phi v^r  \right)
        + v^\phi \partial_\phi v^\phi + \tfrac{1}{r}v^\phi v^r 
        + \tfrac{1}{2}v^z\left( \partial_z v^\phi + \tfrac{1}{r^2}\partial_\phi v^z \right) - \tfrac{1}{3}v^\phi\Theta \\
        \tfrac{1}{2}v^r\left( \partial_r v^z + \partial_z v^r \right) + \tfrac{1}{2}v^\phi\left( \partial_\phi v^z + r^2 \partial_z v^\phi \right) +  v^z\partial_z v^z - \tfrac{1}{3}v^z\Theta
    \end{pmatrix}
\end{align}

The conserved variables and sources are:
\begin{align}
     U &= \begin{pmatrix} 
    \rho \\
    \rho v^r \\
    r^2 \rho v^\phi \\
    \rho v^z \\
    \tfrac{1}{2}\rho v^2 + \rho \epsilon
    \end{pmatrix}, & S &= \begin{pmatrix} 
        0 \\
        r\rho {v^\phi}^2  + \tfrac{1}{r} \left( P + \left(\tfrac{2}{3}\eta - \zeta \right)\Theta\right)  - \tfrac{2}{r}\eta  \left(\partial_\phi v^\phi + \tfrac{1}{r} v^r\right)+  \rho g_r\\
        \rho g_\phi\\
        \rho g_z\\
        \rho \left(v^r g_r + v^\phi g_\phi + v^z g_z\right)
        \end{pmatrix}.
\end{align}

And the fluxes:
\begin{align}
    F^{\hat{r}} = F^r = &= \begin{pmatrix}
        \rho v^r \\
        \rho {v^r}^2 + P + \left(\tfrac{2}{3}\eta - \zeta\right)\Theta - 2\eta \partial_r v^r \\
        r^2\rho v^\phi v^r - \eta \left(\partial_\phi v^r + r^2 \partial_r v^\phi\right) \\
        \rho v^z v^r - \eta \left(\partial_z v^r + \partial_r v^z\right)\\
        \left(\tfrac{1}{2} \rho v^2 + \rho \epsilon + P + \left(\tfrac{2}{3} \eta - \zeta\right) \Theta\right)v^r - 2 \eta \left( {\bf v} \cdot {\bf \sigma}\right)^r
    \end{pmatrix} ,\\
    F^{\hat{\phi}} = rF^{\phi} &= \begin{pmatrix} 
        r \rho v^\phi \\
        r \rho v^r v^\phi - r \eta \left( \partial_r v^\phi + \tfrac{1}{r^2}\partial_\phi v^r\right) \\
        r^3\rho {v^\phi}^2 + rP + r\left(\tfrac{2}{3}\eta - \zeta\right) \Theta - 2 r \eta \left(\partial_\phi v^\phi + \tfrac{1}{r} \partial_r v^r\right) \\
        r\rho v^z v^\phi - r\eta \left(\partial_z v^\phi + \tfrac{1}{r^2}\partial_\phi v^z\right)\\
        r\left(\tfrac{1}{2} \rho v^2 + \rho \epsilon + P + \left(\tfrac{2}{3} \eta - \zeta\right) \Theta\right)v^\phi - 2 r \eta \left( {\bf v} \cdot {\bf \sigma}\right)^\phi
    \end{pmatrix}, \\
    F^{\hat{z}} = F^z &= \begin{pmatrix}
        \rho v^z \\
        \rho v^r v^z - \eta \left( \partial_r v^z + \partial_z v^r\right) \\
        r^2 \rho v^\phi v^z - \eta \left(\partial_\phi v^z + r^2 \partial_z v^\phi\right)\\
        \rho{v^z}^2 + P + \left(\tfrac{2}{3}\eta - \zeta\right)\Theta - 2\eta \partial_z v^z \\
        \left(\tfrac{1}{2} \rho v^2 + \rho \epsilon + P + \left(\tfrac{2}{3} \eta - \zeta\right) \Theta\right)v^z - 2 \eta \left( {\bf v} \cdot {\bf \sigma}\right)^z
    \end{pmatrix}.
\end{align}

\section{Spherical Polar Coordinates}

Spherical polar coordinates $(R, \theta, \phi)$ have a metric tensor $
\gamma_{ij} = \mathrm{diag}(1, R^2, R^2\sin^2\theta)$.  Radial and tensor components transform as in the Cartesian case, latitudinal components obey $v_\theta = R^2 v^\theta$, $v^{\hat{\theta}} = v_{\hat{\theta}} = R v^\theta$, and azimuthal components obey $v_\phi = R^2 \sin^2\theta v^\phi$, $v^{\hat{\phi}} = v_{\hat{\phi}} = R \sin \theta v^\phi$.  We give expressions in terms of the contravariant velocity components $v^R$, $v^\theta$, and $v^\phi$ (note $v^\theta$ and $v^\phi$ are angular velocities measured in e.g [rad/s]).  The unit normals in each dimension are:
\begin{align}
    n^{(R)}_i &= \begin{pmatrix} 1 \\ 0 \\ 0 \end{pmatrix} & n^{(\theta)}_i &= \begin{pmatrix} 0 \\R \\ 0 \end{pmatrix} & n^{(phi)}_i = \begin{pmatrix} 0 \\ 0 \\ R\sin \theta\end{pmatrix}
\end{align}

The divergence is:
\begin{align}
    \Theta = \nabla_k v^k = \partial_R v^R + \partial_\theta v^\theta + \partial_\phi v^\phi + \tfrac{2}{R} v^R + \cot\theta\ v^\theta .
\end{align}

The shear tensor in these coordinates is:
\begin{align}
    \sigma_{i\ }^{\ j} &= \begin{pmatrix}
        \partial_R v^R - \tfrac{1}{3}\Theta & \tfrac{1}{2}\left( \partial_R v^\theta + \tfrac{1}{R^2}\partial_\theta v^R \right) & \tfrac{1}{2}\left( \partial_R v^\phi +\tfrac{1}{R^2 \sin^2\theta} \partial_\phi v^R \right)\\
        \tfrac{1}{2}\left( \partial_\theta v^R + R^2 \partial_R v^\theta \right) & \partial_\theta v^\theta + \tfrac{1}{R}v^R - \tfrac{1}{3}\Theta & \tfrac{1}{2}\left( \partial_\theta v^\phi + \tfrac{1}{\sin^2\theta} \partial_\phi v^\theta \right)\\
        \tfrac{1}{2}\left( \partial_\phi v^R + R^2\sin^2\theta\ \partial_R v^\phi \right) & \tfrac{1}{2}\left( \partial_\phi v^\theta + \sin^2\theta\ \partial_\theta v^\phi \right) & \partial_\phi v^\phi  +  \tfrac{1}{R}v^R + \cot\theta\ v^\theta - \tfrac{1}{3}\Theta
    \end{pmatrix}.
\end{align}

Dotted with velocity (for the energy flux):
\begin{align}
    \left( {\bf v} \cdot {\bf \sigma}\right)^j =  v^k\sigma_k^{\ j} &= \begin{pmatrix}
        v^R \partial_R v^R + \tfrac{1}{2}v^\theta\left( \partial_\theta v^R + R^2 \partial_R v^\theta \right) + \tfrac{1}{2}v^\phi\left( \partial_\phi v^R + R^2\sin^2\theta\ \partial_R v^\phi \right) - \tfrac{1}{3}v^R \Theta \\
        \tfrac{1}{2}v^R\left( \partial_R v^\theta + \tfrac{1}{R^2}\partial_\theta v^R \right) + v^\theta \partial_\theta v^\theta +\tfrac{1}{2}v^\phi\left( \partial_\phi v^\theta + \sin^2\theta\ \partial_\theta v^\phi \right) - \tfrac{1}{3}v^\theta\Theta \\
        \tfrac{1}{2}v^R\left( \partial_R v^\phi +\tfrac{1}{R^2 \sin^2\theta} \partial_\phi v^R \right) + \tfrac{1}{2}v^\theta\left( \partial_\theta v^\phi + \tfrac{1}{\sin^2\theta} \partial_\phi v^\theta \right) + v^\phi \partial_\phi v^\phi - \tfrac{1}{3}v^\phi\Theta
    \end{pmatrix}
\end{align}

The conserved variables are: 
\begin{align}
     U &= \begin{pmatrix} 
    \rho \\
    \rho v^R \\
    R^2 \rho v^\theta \\
    R^2 \sin^2\theta\ \rho v^\phi \\
    \tfrac{1}{2}\rho v^2 + \rho \epsilon
    \end{pmatrix}.
\end{align}

The source terms are:
\begin{align}
S &= \begin{pmatrix} 
    0 \\
    R\rho {v^\theta}^2 + R\sin^2\theta\ \rho {v^\phi}^2 + \tfrac{2}{R} \left( P + \left(\tfrac{2}{3}\eta - \zeta \right)\Theta\right)  - \tfrac{2}{R} \eta  \left(\partial_\theta v^\theta + \partial_\phi v^\phi + \tfrac{2}{R} v^R + \cot\theta\ v^\theta \right)+  \rho g_r\\
    R^2\sin\theta \cos\theta\ \rho{v^\phi}^2 + \cot \theta \left( P + \left(\tfrac{2}{3}\eta - \zeta \right)\Theta\right) - 2\cot\theta\ \eta \left(\partial_\phi v^\phi + \tfrac{1}{R}v^R + \cot \theta\ v^\theta\right) + \rho g_\theta\\
    \rho g_\phi\\
    \rho \left(v^r g_r + v^\phi g_\phi + v^z g_z\right)
    \end{pmatrix}.
\end{align}

And the fluxes:
\begin{align}
    F^{\hat{R}} = F^R &= \begin{pmatrix}
        \rho v^R \\
        \rho {v^R}^2 + P + \left(\tfrac{2}{3}\eta - \zeta\right)\Theta - 2\eta \partial_R v^R \\
        R^2 \rho v^\theta v^R - \eta \left(\partial_\theta v^R + R^2 \partial_R v^\theta\right) \\
        R^2\sin^2\theta\ \rho v^\phi v^R - \eta \left(\partial_\phi v^R + R^2 \sin^2\theta\ \partial_R v^\phi\right)\\
        \left(\tfrac{1}{2} \rho v^2 + \rho \epsilon + P + \left(\tfrac{2}{3} \eta - \zeta\right) \Theta\right)v^R - 2 \eta \left( {\bf v} \cdot {\bf \sigma}\right)^R
    \end{pmatrix} ,\\
    F^{\hat{\theta}} = RF^{\theta} &= \begin{pmatrix} 
        R \rho v^\theta \\
        R \rho v^r v^\theta - R \eta \left( \partial_R v^\theta + \tfrac{1}{R^2}\partial_\theta v^R\right) \\
        R^3\rho {v^\theta}^2 + RP + R\left(\tfrac{2}{3}\eta - \zeta\right) \Theta - 2 R \eta \left(\partial_\theta v^\theta + \tfrac{1}{R} \partial_R v^R\right) \\
        R^3\sin^2\theta \rho v^\phi v^\theta - R\eta \left(\partial_\phi v^\theta + \sin^2\theta\ \partial_\theta v^\phi\right)\\
        R\left(\tfrac{1}{2} \rho v^2 + \rho \epsilon + P + \left(\tfrac{2}{3} \eta - \zeta\right) \Theta\right)v^\theta - 2 R \eta \left( {\bf v} \cdot {\bf \sigma}\right)^\theta
    \end{pmatrix}, \\
    F^{\hat{\phi}} = R\sin\theta\ F^\phi &= \begin{pmatrix}
        R\sin\theta\ \rho v^\phi \\
        R\sin\theta\ \rho v^R v^\phi - R\sin\theta\ \eta \left( \partial_R v^\phi + \tfrac{1}{R^2\sin^2\theta} \partial_\phi v^R\right) \\
        R^3\sin\theta \rho v^\theta v^\phi - \eta \left(\partial_\theta v^\phi + \tfrac{1}{\sin^2\theta} \partial_\phi v^\theta\right)\\
        R^3\sin^3\theta \rho{v^\phi}^2 + R\sin\theta\ P + R\sin\theta\ \left(\tfrac{2}{3}\eta - \zeta\right)\Theta - 2R\sin\theta\ \eta \left(\partial_\phi v^\phi + \tfrac{1}{R}v^R + \cot\theta v^\theta\right) \\
        R\sin\theta\ \left(\tfrac{1}{2} \rho v^2 + \rho \epsilon + P + \left(\tfrac{2}{3} \eta - \zeta\right) \Theta\right)v^\phi - 2R\sin\theta\ \eta \left( {\bf v} \cdot {\bf \sigma}\right)^\phi
    \end{pmatrix}.
\end{align}

\end{document}